\documentclass[prb,twocolumn,showpacs,amsmath,amssymb,floatfix,noshowpacs]{revtex4}

\usepackage[dvips]{graphicx}
\usepackage{dcolumn}
\usepackage{bm}
\usepackage{times}

\begin{document}


\title{
Nanotube-Metal Junctions: 2- and 3- Terminal Electrical Transport
}
                                                                                                                 
\author{ San-Huang Ke,$^{1,2}$ Weitao Yang,$^{1}$ and Harold U. Baranger$^{2}$}
                                                                                                                 
\affiliation{
     $^{\rm 1}$Department of Chemistry, Duke University, Durham, NC 27708-0354
\\
     $^{\rm 2}$Department of Physics, Duke University, Durham, NC 27708-0305
}
                                                                                                                 
\date{April 4, 2006; \textbf{J. Chem. Phys. 124, 181102 (2006)}; DOI: 10.1063/1.2200356}
                                                                                                                 
\begin{abstract}
We address the quality of electrical contact between carbon nanotubes and
metallic electrodes by performing first-principles calculations for the
electron transmission through ideal 2- and
3-terminal junctions, thus revealing the physical limit of tube-metal
conduction.
The structural model constructed involves surrounding the tube by the metal
atoms of
the electrode as in most experiments;
we consider metallic (5,5) and \textit{n}-doped semiconducting
(10,0) tubes surrounded by Au or Pd. In the case of metallic tubes, the contact
conductance is shown to approach the ideal $4e^2/h$ in the limit of large
contact area.
For three-terminals, the division of flux among the different transmission
channels depends strongly on the metal material. A Pd electrode has nearly
perfect tube-electrode transmission and therefore turns off the straight
transport along the tube. Our results are in good agreement with some recent
experimental reports and clarify a fundamental discrepancy between theory and
experiment.
\end{abstract}

\maketitle


Carbon nanotubes (CNTs) have received extensive experimental attention for more
than a decade \cite{Avouris04403,Mceuen04272,Heer04281}, and are considered a
possible basis for nanoelectronic technology independent of silicon. A major
issue is the quality of CNT/metal contacts: obtaining the minimum contact
resistance is critical to access the intrinsic electric properties of CNTs.
Despite extensive experimental effort to improve the contact transparency and
reveal the relevant factors behind it -- metal material, contact structure, and
type of tube, for instance -- a clear picture is still not available.

On the theoretical side, it is highly desirable to be able to simulate from the
first principles the
electron transport through CNT/metal junctions and thus to improve our
understanding of this important issue. So far, first-principles 
studies of the contact transparency between metallic
CNTs and metals have been carried out [(3,3), (4,4), or (5,5) tubes with Al,
Au, or Ti, for instance]
\cite{Taylor01245407,Nardelli01245423,Palacios03106801,Liu03193409}. The
results are, however, quite scattered, and agreement between theory and
experiment has not yet been achieved: For Al electrodes, an equilibrium
conductance of $\sim\!1\,G_0$ ($=2e^2/h$, the conductance quanta) was found 
by some calculations \cite{Taylor01245407,Palacios03106801} while
$\sim\!2\,G_0$ was found by another \cite{Nardelli01245423}. 
For Ti electrodes, $1.7\,G_0$ was obtained by one calculation
\cite{Liu03193409} and $\sim\!1.2\,G_0$ by another \cite{Palacios03106801}, 
while experiment \cite{Kong01106801} found $\sim\!2\,G_0$. 
For Au electrodes, a recent calculation \cite{Palacios03106801} showed
$\sim\!1\,G_0$ while a value of $1.5\,G_0$ was found experimentally
\cite{Nygard00342}. 
On the other hand, a model calculation \cite{Choi99R14009} using a jellium model for the electrode
even showed that the ideal CNT/metal conductance will not be larger than
$1\,G_0$. 
With regard to the metal used, it was found theoretically that Ti is better
than either Au or Al for contact transparency \cite{Palacios03106801}.  On the
other hand, experimentally, Pd, for which no theoretical calculation is
available, was found superior to Ti \cite{Mann031541}. 

It has been unclear how to explain these discrepancies.
One possible reason, as is supported by this work, 
is the difference in contact structure between theory and
experiment. In all these calculations, simplified contact models were adopted
due to computational cost or methodology: On the carbon side, the tube-metal
connection is made by either straight $\sigma$-bonds from the tube end or
$\pi$-bonds from the side of the tube, while the metal is modeled by a thin
nanowire or small cluster. However, in most experimental situations, a tube is
surrounded by metal atoms \cite{Soh99627,Martel01159,McEuen0278,
Wind03058301,Yaish04046401,Biercuk041}. 
As the contact structure and quality changes from case
to case, experimental results are also scattered. In such a situation, the key
contribution to be made by theory is to reveal the physical limit to which
experimental measurement may approach by improving the contact quality. 

Furthermore, compared to metallic CNTs, semiconducting CNTs are much more
important for potential electronic applications. This naturally raises the
issue of contact transparency between doped semiconducting tubes and metal
junctions. In addition, electron transport through multi-terminal structures is
a key property in moving toward applications. Neither of these fundamental
issues has been tackled previously, to the best of our knowledge, using
first-principles calculations.  

\begin{figure}[b]
\includegraphics[angle= 0,width=8.0cm]{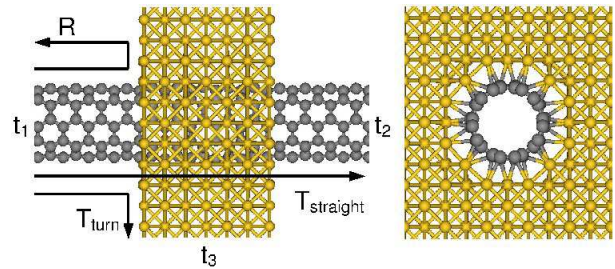}
\caption{Side and top views of the device region of a (5,5)-tube-Au junction,
where the Au electrode consists of 7 atomic layers (7L). 
The length of the tube in the device region is 32.6{\AA} and 
the dimensions in the lateral directions are 20.4 and 22.44{\AA},
respectively.
The three terminals are denoted by t$_1$, t$_2$, and t$_3$.
}
\label{fig_str}
\end{figure}

\begin{figure*}[t]
\includegraphics[angle= 0,width=11cm]{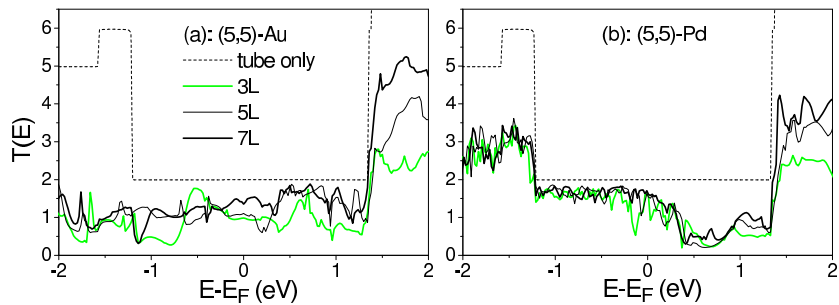}
\caption{Two-terminal transmission $T_{\rm 2term.}(E)$ (tube to metal) for (a) (5,5)-Au 
and (b) (5,5)-Pd with different widths of the metal electrode, as indicated in
the legends. Also plotted is the transmission for a pure (5,5) tube, as a
comparison.} 
\label{fig_t2t5x5}
\end{figure*}

In this paper, we first construct a better
\cite{Soh99627,Martel01159,McEuen0278,Wind03058301,Yaish04046401,Biercuk041}
structural model of an ideal contact, as shown in Fig.~\ref{fig_str} for the
device region, in which
an infinitely long tube is surrounded by metal atoms of an electrode.  
We use this structural model, first, to investigate the two-terminal tube-metal
transparency, $T_{\rm 2term.}$ (from terminal 1 to 3), obtaining the physical limit of
tube-metal conductance for an ideal contact. Second, we then study
three-terminal electron transport by including explicitly the self-energies of
all the leads shown in Fig.~\ref{fig_str}. Specifically, we investigate the
division of current among transmission straight-through the junction,
probability to turn the corner, and reflection. In both 2- and 3- terminal
cases, we consider the dependence on the contact quality (simulated by the
width of the metal electrode), the metal material, and the type of tube
(metallic or semiconducting). 


The structural model (Fig.~\ref{fig_str}) can be regarded as an ideal CNT-metal
junction because of three features: (1)~The tube is perfect and infinitely
long, thus eliminating the quantization in the transport direction which
bedevils calculations \cite{Palacios03106801} using a short CNT segment. 
(2)~The tube is surrounded by metal atoms, as is usual in experiments
\cite{Soh99627,Martel01159,McEuen0278,Wind03058301,Yaish04046401,Biercuk041}.
(3)~The large active region around the countact
ensures that the tube-metal interaction is fully included. 

Contact quality will depend experimentally on the number of good carbon-metal
connections, which is essentially determined by whether the metal wets the CNT
surface. If the metal (like Au) does not wet the CNT surface, there will be few
connections even though the electrode is large.  In our mod\-els, because of
the ideal structure, the number of good car\-bon-metal connections is
substantial even for a small electrode (Fig.~\ref{fig_str}, top view). As a
proxy, we change the number of good carbon-metal connections by adjusting the
contact width, in this way simulating the changing contact
quality. 

We consider three widths for the metal electrodes -- 3, 5, and 7 atomic layers
(denoted by 3L, 5L, and 7L) -- and two kinds of metals -- Au and Pd. 
The electronic states of the two electrodes are very different: Au has active \textit{s}
states while Pd has only \textit{d} states. We consider three kinds of CNTs:
(5,5), (10,0), and Na-doped (10,0) -- denoted Na@(10,0) hereafter -- in which
dopant Na atoms are adsorbed periodically on the inner surface of the tube. 
[The adsorption position is determined by minimizing the atomic force on the
Na atom and is found to be slightly off the tube axis and 
approximately above one of the C atoms.] 
The concentration is one Na atom per 40 C atoms. Because of the large charge
transfer from Na to the tube ($\sim\!0.7$ electrons per Na atom by Mulliken
population analysis), the (10,0) tube is therefore heavily \textit{n}-doped. 

The electronic structure of the junctions is calculated by density functional
theory (DFT) using periodic boundary conditions and a localized basis set with 
a finite range \cite{Soler022745}. 
In the device region, as shown in Fig.~\ref{fig_str}, the number
of atoms included is $\sim$ 350 -- 500, depending on the
metal width and material.
The large lateral dimensions (see Fig.~\ref{fig_str} (top view)) ensure 
that the separation among the tube and its
images is larger than 12{\AA} which is much larger than the range of the basis
functions used. As a result, a good convergence with respect to the lateral
dimensions can be expected. The convergence with respect to the length of the
tube in the device region is also found good, as will be discussed later.

We adopt a DFT supercell which is larger than the
device region shown in Fig.~\ref{fig_str} for all the different systems: $\sim$
500 -- 700 atoms are included, depending on the electrode width and material. 
To avoid too expensive computational cost,  
a single-zeta plus polarization basis set (SZP) is adopted
for all atomic species. Our test calculation for a small system shows that the
use of SZP results in only minor differences from results using a higher-level
double-zeta plus polarization basis set. We use optimized Troullier-Martins
pseudopotentials \cite{Troullier911993} for the atomic cores and the PBE
version of the generalized gradient approximation \cite{Perdew963865} for the
electron exchange and correlation. The contact atomic structure is optimized by
eliminating the atomic forces on the atoms around the contact region, i.e., 
the carbon atoms underneath the metal and the metal atoms contacting the tube. 
The relaxation is found quite small (see Fig.~\ref{fig_str} (top view) for 
the (5,5)-Au system) because of the choices of the tubes, (5,5) and (10,0), and
the orientation of the metals, (001).

Electron transport through the CNT-metal junction is calculated using a 
Green function method \cite{Datta95,Ke04085410}, in which the device region and
the leads (not shown in Fig.~\ref{fig_str}) are treated exactly on the same
footing \cite{Ke04085410}.  The retarded Green function of the
device region, $\mathbf{G}_{D}(E)$, is determined by the Hamiltonian given by
DFT [$\mathbf{H}_{D}$] combined with the self-energies for the semi-infinite
leads [${\Sigma}_{i}(E)$]:
\begin{equation}
\mathbf{G}_{D}(E) = \Big[ E\mathbf{S}_{D} - \mathbf{H}_{D}
  -\sum_{i=1}^{N_t} {\Sigma}_{i}(E) \Big]^{-1}, \label{equ_g}
\end{equation}
where $N_t$ is the number of leads or terminals. The transmission at any
energy, $T(E)$, is calculated from the Green function, and the conductance,
$G$, then follows from a Landauer-type relation.  We adopt the following
notation: 
$T_{\rm 2term.}$ is the two-terminal transmission between tube and metal (terminals 1
and 3) \textit{without} electrode 2 attached, which means that
the tube on the right side of Fig.1 (side view) is somehow terminated and all the
electron wave entering that part of tube will be totally reflected.  
$T_{\rm straight}$ is the transmission
straight through the junction from tube to tube, $T_{\rm turn}$ is the tube-to-metal
transmission \textit{with} the presence of electrode 2, and $R$ is reflection
from the tube back to itself. 


\begin{figure*}[t]
\includegraphics[angle= 0,width=11.6cm]{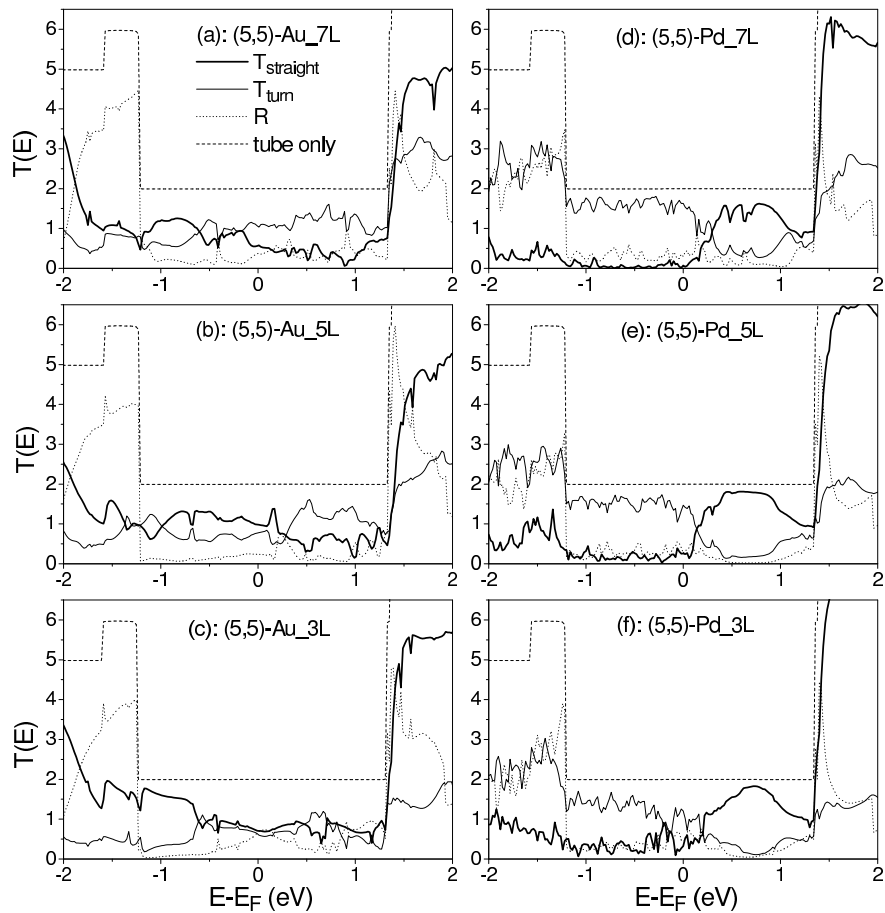}
\caption{Three-terminal transmission functions $T_{\rm straight}(E)$,
$T_{\rm turn}(E)$, and $R(E)$
for (5,5)-Au [first column, (a)-(c)] and (5,5)-Pd [second column, (d)-(f)]. The
width of the metal electrode is indicated in the legends. Also plotted is the
transmission function of a pure (5,5) tube as a comparison (dashed). The Pd
traces converge much more quickly than those for Au.}
\label{fig_t3t5x5}
\end{figure*}

In Fig.~\ref{fig_t2t5x5} we show the transmission $T_{\rm 2term.}(E)$ for the (5,5)-Au
and (5,5)-Pd junctions with different widths of the metal electrodes. As a
comparison, $T(E)$ for the pure (5,5) tube is also plotted; it shows, as
expected, perfect steps and two channels around the Fermi energy. Note that
throughout this work the position of the Fermi energy is set to be exactly the
same as in the pure tube. This is required since very far from the device
region the tube (lead) is completely charge neutral. 

In Fig.~\ref{fig_t2t5x5}, the dependence of $T_{\rm 2term.}(E)$ on the electrode width
for the two metals is clear:  For Au, the dependence is strong. In contrast,
for Pd the dependence is much weaker: the overall shape of $T_{\rm 2term.}(E)$ for the
three widths is similar, and the result is nearly the same for 5L and 7L.
This result for Pd also indicates that in our calculation the convergence with 
respect to the length of the tube included in the device region is good.
Despite the difference in width dependence, the equilibrium conductance, $G$,
of both systems increases with increasing electrode width: for the thinnest,
hence poor quality, electrode $G \!\sim\! G_0$, while it approaches the
physical limit of $2\,G_0$ for the widest electrode. 

Even the widest electrode (7L) considered here is, of course, still much
thinner than those used in experiments. However, our calculation bears on real
experimental situations because the number of good carbon-metal connections may
be similar in that the contact structure considered here is perfect while in
real experimental situations it is usually far from perfect. For example, Au is
thought not to wet the tube surface but rather will form nanoparticles near the
surface \cite{JieLiu}. As a result, although the Au electrode can be very wide,
the tube will pass through the space between these Au nanoparticles with very
few good C-Au connections formed. In this case increasing the diameter of the
tube will improve the contact quality as more C-Au connections will be formed
and, therefore, increase the conductance. In fact, a recent experiment using
tubes with different diameters \cite{Javey03654} shows that a larger diameter
yields a larger conductance. Obviously, increasing the electrode width here is
similar to increasing the tube diameter: both increase the number of C-Au
connections and so increase the conductance. \textit{A key result here is that
for good contact quality, the equilibrium conductance approaches the physical
limit of 2 $G_0$ for both Au and Pd.} This finding differs from a previous
first-principles calculation \cite{Palacios03106801} which used a short segment of tube
contacted by two Au electrodes through several carbon-$\pi$ bonds, and also
differs from a previous model calculation \cite{Choi99R14009} which shows that CNT/metal
conductance will not be larger than 1 $G_0$, but is
consistent with recent experiments \cite{Kong01106801,Mann031541}. 


\begin{figure*}[t]
\includegraphics[angle= 0,width=12cm]{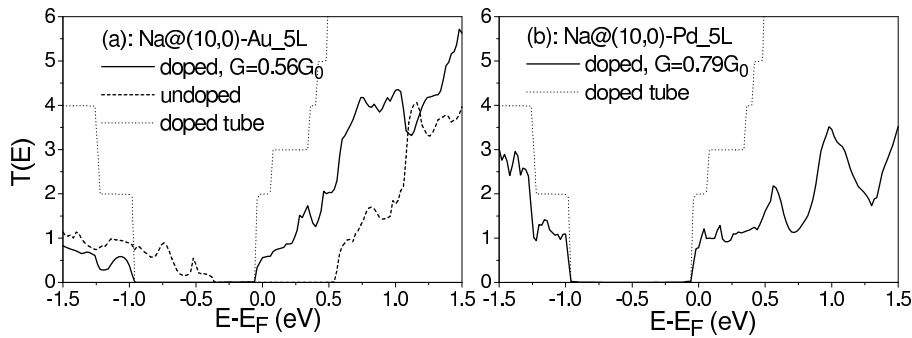}
\caption{Two-terminal transmission, $T_{\rm 2term.}(E)$, with semiconducting tubes: (a)
Na@(10,0)-Au and (b) Na@(10,0)-Pd systems with an electrode width of 5L.  
The transmission function of the bare Na@(10,0) tube is shown in each panel for
comparison; in addition, (a) includes $T_{\rm 2term.}(E)$ for the undoped (10,0)-Au
system.} 
\label{fig_t2t10x0}
\end{figure*}

The three-terminal transmission in the (5,5)-metal case is shown in
Fig.~\ref{fig_t3t5x5} for the different metal widths. [Note that reflection
$R(E) \!=\! T_{\rm tube}(E) \!-\! T_{\rm straight}(E) \!-\! T_{\rm turn}(E)$.] 
For the widest case (7L) (see Figs.~\ref{fig_t3t5x5} (a) and (d)) a striking feature
of the result for Pd is that the straight-through transmission is almost turned
off while transmission into the  
metal electrode dominates. This feature is in very good agreement with a recent
experiment \cite{Mann031541} showing that a Pd electrode suppresses inner
electron transport through the tube. For Au, $T_{\rm turn}$ is also larger
than $T_{\rm straight}$ but they are still comparable and the latter is not turned
off.

In Figs.~\ref{fig_t3t5x5}, we see that for Au the
apportioning of transmission among the different components depends
significantly on the electrode width: For small width, $T_{\rm straight}(E)$
and $T_{\rm turn}(E)$
are comparable around the Fermi energy, while for the widest electrode (7L),
$T_{\rm turn}(E)$ begins to dominate due to the improved contact quality. In the Pd
system, the convergence as a function of width is much faster (as for the
two-terminal case Fig.~\ref{fig_t2t5x5}); indeed, one sees only minor changes
between the curves. 
This quick convergence indicates that the carrier injection takes place mainly 
at the very edge of the junction, being consistent with the experimental
observation \cite{Mann031541}.
Thus for Pd, the contact quality is already good (i.e., the width is large
enough) even
though the junction is very thin. Since the contact structure is the same, this
difference between Au and Pd electrodes is due to their different electronic
states: The Pd \textit{d} states have stronger interaction with C \textit{p}
states than Au \textit{s-d} states. As a result, Pd is a better electrode
material than Au. 


Two additional features of Figs.~\ref{fig_t3t5x5} are
worth comments. First, $T_{\rm turn}(E)$ is always smaller than the two-terminal result
$T_{\rm 2term.}(E)$ of Fig.~\ref{fig_t2t5x5}. This is reasonable since in the
three-terminal case some of the initial flux escapes into electrode 2 while for
two terminals that flux will be totally reflected. Second, the reflection
increases in all cases near the energies where more modes start to propagate in
the pure tube. These are examples of threshold singularities
\cite{LandauLifshitzQM}.  


We now turn from metallic tubes to discussing the doped (10,0) semiconducting
tube. The doping is achieved by adding Na atoms periodically on the inner
surface of the (10,0) tube with a high concentration as described earlier. 

The two terminal transmission is shown in Fig.~\ref{fig_t2t10x0} for both Au
and Pd electrodes. For comparison, the transmission is shown for two additional
cases: $T_{\rm 2term.}(E)$ for the Na@(10,0) tube alone and for the undoped (10,0)-Au
system. For the undoped (10,0)-Au system, the Fermi energy is at the middle of
the gap because one of the leads is the semiconducting (10,0) tube.  After the
tube is \textit{n}-doped, its Fermi energy enters the conduction band, and the
transmission through the Na@(10,0) tube in the absence of any metal has a value
of 2 around the Fermi energy.  As in the metallic case, here the Pd electrode
results in a larger two-terminal conductance ($0.79\,G_0$) than the Au
electrode ($0.56\,G_0$). Both are smaller than the two-terminal conductance of
the (5,5) systems. 

An interesting feature of the results in Fig.~\ref{fig_t2t10x0} is that
$T_{\rm 2term.}(E)$ for the doped Na@(10,0)-Au case is approximately a constant shift of
that for the undoped (10,0)-Au system.  Thus the role of the dopants is mainly
to give electrons to the tube, and therefore to shift its Fermi energy, without
noticeably altering the electronic structure of the tube. Our calculations
indicate this is a general rule of thumb. Following this idea, we can
approximately get the two-terminal conductance of a $p$-doped (10,0)-metal
junction by simply shifting the Fermi energy slightly into the valence band (to
around $-1$\,eV in Fig.~\ref{fig_t2t10x0}.  For a Pd electrode, this $p$-type
conductance is about $1\,G_0$ which is in good agreement with a recent
experimental report where the Fermi energy of a semiconducting CNT contacted by
Pd electrodes was shifted into the valence band by applying a back gate [see
Fig. 1 (c) in Ref.~\cite{Javey03654}]. 

\begin{figure*}[t]
\includegraphics[angle= 0,width=12cm]{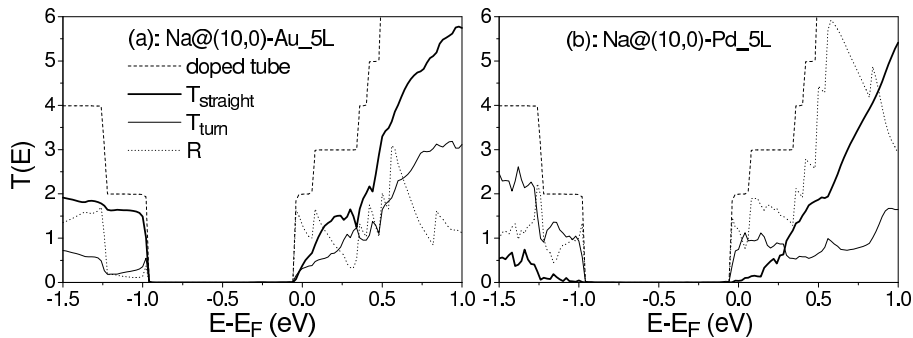}
\caption{Three-terminal case with semiconducting tubes: $T_{\rm straight}(E)$,
$T_{\rm turn}(E)$,
and $R(E)$ for (a) Na@(10,0)-Au and (b) Na@(10,0)-Pd with an electrode width of
5L. For comparison, the transmission function of the Na@(10,0) tube is also
shown.} 
\label{fig_t3t10x0}
\end{figure*}

The three-terminal transmission functions for the semiconducting tubes cases
are shown in Fig.~\ref{fig_t3t10x0}. For Na@(10,0)-Au, the straight-through and
tube-metal transmission around the Fermi energy are comparable, just like for
the corresponding (5,5) system [Fig.~\ref{fig_t3t5x5} (b)]. Both of them are,
however, much smaller than the reflection because the Fermi energy now is at
the edge of the gap.  For a Pd electrode, on the other hand, $T_{\rm turn}$ is
comparable to the reflection, near the Fermi energy, and both of them are much
larger then the straight-through transport. Therefore, for an $n$-doped
semiconducting (10,0) tube, the Pd electrode turns off conduction through the
tube, just like for the metallic (5,5) tube. 

\newpage
In summary, we have calculated transmission through both 2- and 3- terminal
CNT-metal junctions using an ideal structure in which the tube is surrounded by
a Au or Pd electrode. In this way we have established the physical limit to
which experiments will approach by improving the contact quality. 

The main quantities studied are the contact transparency in the case of
two-terminal systems and the division among the different transmission
coefficients for three-terminals. 
The two main findings are:

(1) For two-terminal junctions, when the tube is metallic the conductance will
approach $2\,G_0$ as the contact quality improves, while for the doped
semiconducting tube systems, the conductance is about $1\,G_0$. In both cases,
Pd is better than Au for contact transparency. 

(2) For three terminal junctions, the relative magnitude of the different types
of transmission depends significantly on the metal material. The ``better'' Pd
electrode yields near perfect transmission between the tube and electrode and,
therefore, turns off the straight transport through the tube for both the (5,5)
and Na@(10,0) tubes. 

Our results are in good agreement with recent experiments 
and clarify a fundamental discrepancy between theory and
experiment.

{\bf Acknowledgments.}
This work was supported in part by the NSF (DMR-0506953).

\end{document}